\documentclass[floatfix,notitlepage,prd,reprint,superscriptaddress,twocolumn]{revtex4-1}
\usepackage{graphicx,amsmath,bm,natbib,color}

\begin{document}

%opening
\title{Splashback in accreting dark matter halos}
\author{Susmita Adhikari}
\affiliation{Department of Astronomy, University of Illinois at Urbana-Champaign}
\author{Neal Dalal}
\affiliation{Department of Astronomy, University of Illinois at Urbana-Champaign}
\affiliation{Department of Physics, University of Illinois at Urbana-Champaign}
\author{Robert T. Chamberlain}
\affiliation{Department of Physics, University of Illinois at Urbana-Champaign}

\begin{abstract}
Recent work has shown that density profiles in the outskirts of dark matter halos can become extremely steep over a narrow range of radius.  This behavior is produced by splashback material on its first apocentric passage after accretion.  We show that the location of this splashback feature may be understood quite simply, from first principles.  We present a simple model, based on spherical collapse, that accurately predicts the location of splashback without any free parameters.  The important quantities that determine the splashback radius are accretion rate and redshift.  
\end{abstract}

\maketitle

\section{Introduction} \label{intro}

The structure of dark matter halos has attracted both theoretical and observational interest over several decades.  Beginning with the pioneering work of \citet{GunnGott72}, numerous papers have investigated the formation of bound virialized structures that gravitationally collapse in an expanding universe.  A key breakthrough was provided by \citet{FG84}, who studied the self-similar collapse of scale-free perturbations, and identified several key physical processes that determine halo profiles.  As numerical simulations of halo formation have progressed \citep{ViaLactea1,ViaLactea2,GHalo,Aquarius}, producing increasingly precise calculations of halo structure, the ideas presented in \citep{FG84} have proven fundamental towards understanding the simulation results \citep{LD10,Dalal2010}.  

Much of this theoretical work has focused on the interior structure of halos, while the outer profiles of halos have received somewhat less attention.  Recently, however, \citet{Diemer2014} studied the outskirts of simulated halos and discovered that the outer density profiles of halos exhibit steep logarithmic slopes, $d(\log\rho)/d(\log r) \lesssim -4$, over a narrow range of radii.  This behavior is inconsistent with standard fitting functions used to characterize halo shapes, like the NFW profile \citep{NFW96,NFW97} or the Einasto profile \citep{Merritt05,Merritt06}.  \citeauthor{Diemer2014} found that the location of this sharp feature depends principally on the accretion rate $s=d(\log M)/d(\log a)$ of the halos, and they provided a fitting function for its location as a function of $s$.

As argued by \citet{Diemer2014}, the local steepening that they observed is produced by a caustic, associated with the splashback of material that has been recently accreted by the halo.  Caustics arise when the orbits of different particles pile up at similar locations, frequently near the apocenters of the orbits.  For example, the spherically symmetric similarity solutions of \citet{FG84} exhibit pronounced caustics.  Similar features arise in 3D similarity solutions of the collapse of triaxial peaks \citep{LD10}, although the features are not as prominent in the spherically averaged density profile, because the caustics are not spherically symmetric in general (see \S\ref{sec:sim}).  In dark matter halos from cosmological N-body simulations, the density enhancements associated with caustics are difficult to detect, not only because of triaxiality, but also because of the effects of small-scale substructure, both of which act to smear out caustics spatially \citep{Diemand2008,Vogelsberger2011}.  In real galaxies, these caustics are observed as radial shells with sharp edges \citep{Cooper2011}.  

\begin{figure}
\includegraphics[width=0.6\textwidth, trim=1.5in 0.7in 1in 0.25in]{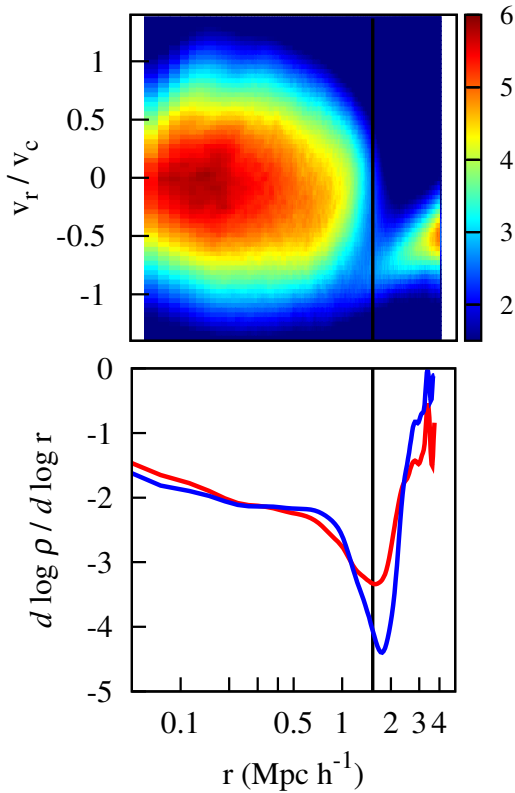}
\caption{Top: The phase space diagram for halos from the MDR1 simulation in the mass range $M=1-4\times10^{14} h^{-1} M_\odot$.  The colorbar shows the number of particles within each phase space pixel.  The pixel spacing is linear in both $r$ and $v$, so the number is proportional to $r^2\rho$.  Bottom:  The local slope of the density of all particles (red) and particles with $|v_r|<0.4\,v_{\rm circ}$ (blue), as a function of radius $r$.  The location of the feature in the local slope coincides with the outer caustic at the splashback radius.}  
\label{phasespace}
\end{figure}

As noted above, radial caustics are associated with the pileup of orbits near apoapse.  The outermost caustic is associated with the first apoapse after collapse, termed splashback.  Figure \ref{phasespace} illustrates that the local steepening discussed by \citep{Diemer2014} coincides with the splashback radius.  The figure plots the phase space structure of the particles near dark matter halos taken from the publicly available MultiDark Simulation\footnote{http://www.multidark.org/MultiDark}, along with the radial dependence of the local logarithmic slope of the density $d(\log\rho)/d(\log r)$.  The location where $d(\log\rho)/d(\log r) < -3$ coincides with splashback, the outermost radius attained by particles following their collapse into halos.  As the phase-space diagram illustrates, the splashback radius is near the location of a radial caustic, where the slope of the phase space sheet becomes vertical.  To further illustrate this point, the figure plots the density slope of only the particles near splashback, i.e.\ those with $|v_r| < 0.4\,v_{\rm circ}$.  Among the particles near splashback, the steepening of the density slope becomes even more pronounced.

The steepening feature in the outer profile is therefore determined by the splashback radius of recently accreted material. Since splashback occurs only half an orbit after collapse, a relatively simple treatment of the orbital dynamics should suffice to capture the physics setting the splashback radius.  In this paper, we show that this is indeed the case.  We construct an extremely simple model for splashback, based largely on the spherical collapse model of \citep{GunnGott72}.  We then compare the predictions of our model with N-body simulations, and show that it accurately predicts the location of the steepening feature for a variety of halos with different mass, redshift, and accretion rate.

\section{Toy model for splashback}
\label{toy}

As noted above, the steepening feature occurs near the splashback radius.  To predict the location of this feature, we therefore must predict where splashback occurs following collapse.  One estimate for the location of the splashback radius may be constructed from the spherical collapse model \citep{GunnGott72}.  This model computes the nonlinear evolution of a spherical shell, assuming that the mass interior to the shell is overdense, and that the interior mass is a constant (i.e., the model neglects shell crossing).  The equation of motion is therefore quite simple, 
\begin{equation}
\ddot r = -\frac{GM}{r^2} + \frac{\Lambda\,c^2}{3} r,
\label{EOM}\end{equation}
where $M$ is a constant.  For $\Omega_M=1$ and $\Lambda=0$, this model predicts that turnaround occurs at the time when the linearly evolved density reaches $\delta_l\approx 1.06$, and collapse to $r=0$ occurs at the time when $\delta_l\approx 1.69$.  Once the infalling shell enters the virialized region, the assumption of constant interior mass becomes invalid.  Estimating virialization as the time when $2\,{\rm KE}+{\rm PE}=\dot r^2 - GM/r=0$ gives $r_{\rm vir}\approx r_{\rm ta}/2$, corresponding to a nonlinear overdensity of $\Delta_{\rm vir}=18\pi^2$.  For $\Omega_M\neq 1$, these expressions are somewhat modified \citep{Eke1996}.

Following entry into the halo, the shell begins to orbit in the halo potential.  If we continue to assume spherical symmetry, then we can still use Eqn.\ \eqref{EOM} to compute the motion; the only change is that now the mass $M$ interior to the shell is not constant, but instead depends on radius $r$ and time $t$.  The time dependence of the mass profile is determined by the accretion rate of the halo.  Let us suppose that the halo mass grows as $M_{\rm tot} \propto a^s$, where $s=d(\log M_{\rm tot})/d(\log a)$.  Then the halo radius scales as $R\propto a^{1+s/3}$.  We assume that the halo mass distribution is given by the NFW profile \citep{NFW97}, 
\begin{equation}
M(r) = M_{\rm tot} \frac{f_{\rm NFW}(r/r_s)}{f_{\rm NFW}(R/r_s)},
\label{mass}
\end{equation}
where $r_s$ is the NFW scale radius, and $f_{\rm NFW}(x) = \log(1+x)-x/(1+x)$.  The NFW concentration $c\equiv R/r_s$ sets the slope of the mass profile at the halo boundary, and we choose $c$ such that the outer slope is given by $d\log M/d\log r=3s/(3+s)$ at $r=R$ \citep{FG84,LD10,Dalal2010}. 

\begin{figure}[t]
\centerline{\includegraphics[width=0.60\textwidth]{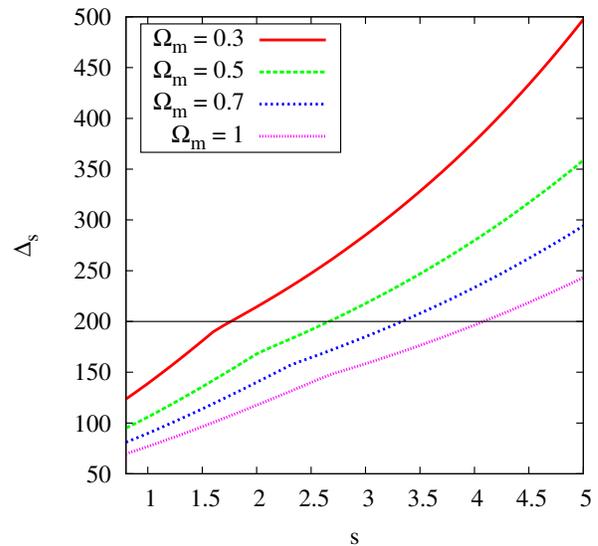}}
\caption{Overdensity enclosed within the splashback radius as a function of accretion rate $s=d\log M/d\log a$.  Large $\Delta$ corresponds to small radius within the halo.   For reference, the horizontal line at $\Delta=200$ indicates the usual definition of the halo boundary.  Halos with high accretion rate have a larger enclosed overdensity within the splashback radius than halos with low accretion rates, corresponding to a smaller splashback radius for rapidly accreting halos.  Additionally, the enclosed overdensity increases as $\Omega_m$ becomes smaller, meaning that at low redshift, the splashback occurs at a smaller fraction of $r_{200m}$.}
\label{Delta_s}
\end{figure}

Equations \eqref{EOM}-\eqref{mass} fully specify our model.  We use spherical collapse with constant enclosed mass $M$ until the shell radius reaches $r_i=r_{\rm ta}/2$.  Thereafter, we assume the mass profile is given by Eqn.\ \eqref{mass}. We integrate the motion of the shell in this potential until splashback, when $\dot r=0$.  From the radius and time of splashback, $r_s$ and $t_s$, we can determine the enclosed mass, the enclosed density, and the background mean density, allowing us to determine the enclosed overdensity inside the splashback radius, $\Delta_s$.  Our results do not strongly depend on our assumed mass profile inside the halo.  For example, using an isothermal profile instead of NFW gives results that are consistent at the $\sim 10\%$ level.  Figure \ref{Delta_s} shows the predicted values of the enclosed overdensity. Throughout this paper, we define overdensities relative to the mean matter density, not the critical density.  In our model, $\Delta_s$ depends only on the halo's accretion rate $s$, along with the values of the background cosmological parameters $\Omega_M$ and $\Omega_\Lambda$ at the time the halo is observed.  The behavior we find is unsurprising.  As the accretion rate is increased (larger $s$), the potential deepens more quickly in time, resulting in splashback occuring at a smaller radius, or equivalently, at a larger enclosed overdensity $\Delta_s$.  Similarly, at low redshift when $\Omega_M$ diminishes and $\Omega_\Lambda$ increases, the mean background density of the universe $\bar\rho_m$ decreases more during the time between turnaround and splashback, again resulting in a larger $\Delta_s$.

Finally, although the model presented here is extremely simple to evaluate, we also provide a very rough fitting function to approximate the location of splashback:
\begin{equation}
\Delta_s \approx A\,\Omega_M^{-b-c\, s} e^{d\Omega_M+e\,s^{3/4}},
\end{equation}
with fitted parameters $A=38$, $b=0.57$, $c=0.02$, $d=0.2$, $e=0.52$.  This fitting function has $\sim 5\%$ accuracy over the range $0.5<s<4$, $0.1<\Omega_M<1$.  The results shown in this paper do not use this fitting function, since it is equally simple to evaluate the spherical toy model.

\section{Comparison with simulations}
\label{sec:sim}

\begin{figure}

\centerline{\includegraphics[width=0.5\textwidth, trim = 1.5cm 0 1.5cm 0,clip]{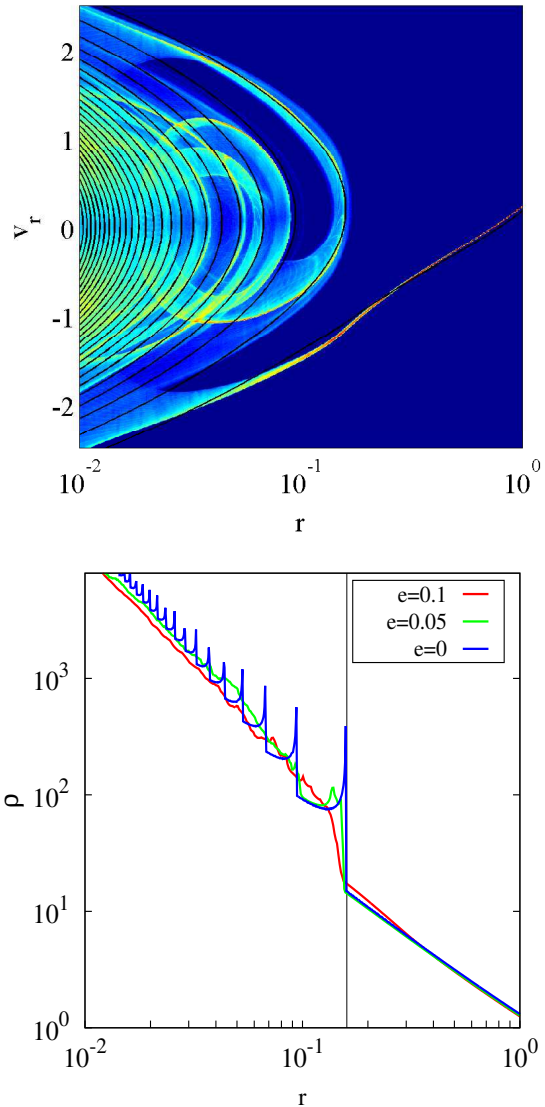}}
\caption{Caustics for self-similar halos \citep{FG84,LD10} with accretion rate $s=3$.  The top panel shows the phase space diagram for spherically symmetric collapse (solid black curve) and for 3D collapse with $e=0.05$ (colormap), while the bottom panel shows the  density vs.\ radius.  The vertical line in the bottom panel indicates the splashback radius predicted by the spherical collapse model for this value of $s$.  As the density profiles demonstrate, the caustic location depends mainly on accretion rate, with little if any dependence on the initial ellipticity $e$.  However, the caustic width does depend on $e$, apparently because the shape of the splashback surface is related to the initial ellipticity.  
\label{ssim}}
\end{figure}

In this section, we compare the predictions of the toy model described in the previous section with results of numerical simulations.  First, we compare our model with the similarity solutions that arise from the collapse of scale-free perturbations \citep{FG84,LD10}.  Fig.\ \ref{ssim} shows one example, for accretion rate $s=3$.  In all cases, we find good agreement between the caustic location obtained in the similarity solution and that predicted by the toy model.  This even holds true for collapse of highly triaxial perturbations: the main effect of the triaxiality is to make the splashback surface nonspherical, reducing the maximal depth of the slope of the spherically averaged profile, while preserving the mean radial location of splashback.  

%\begin{figure}
%\includegraphics[width=0.45\textwidth,trim = 1.5cm 0 2.5cm 0]{ssimph_lowres.eps}
%\centerline{\includegraphics[width=0.59\textwidth, trim = 0 0 0 0]{ssma.eps}}
%\caption{Caustics for self-similar halos \citep{FG84,LD10} with accretion rate $s=3$.  The top %panel shows the phase space diagram for spherically symmetric collapse (black line) and for 3D %collapse with $e=0.05$ (colormap), while the bottom panel shows the  density vs.\ radius.  The %vertical line in the bottom panel indicates the splashback radius predicted by the spherical %collapse model for this value of $s$.  As the density profiles demonstrate, the caustic location %depends mainly on accretion rate, with little if any dependence on the initial ellipticity $e$.  %However, the caustic {\em depth} does depend on $e$, apparently because the shape of the splashback %surface is related to the initial ellipticity.  
%\label{ssim}}
%\end{figure}

\begin{figure}
\centerline{\includegraphics[width=0.65\textwidth]{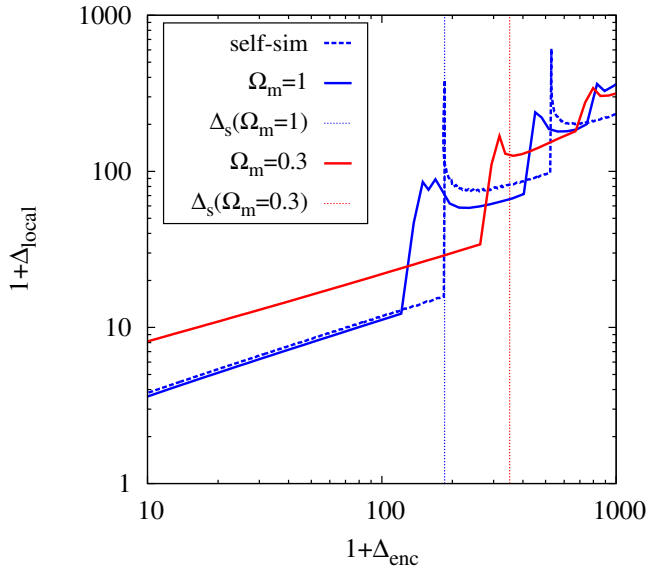}}%{s3.eps}}
\caption{Halo profiles for 1-D spherical collapse with accretion rate $s=3$. The x-axis shows the averaged enclosed overdensity inside each radius, while the y-axis shows the local overdensity at that radius.  The blue curves correspond to $\Omega_M=1$.  The solid curve shows the profile obtained from the 1-D N-body simulation, while the dashed curve shows the similarity solution.  The red curve shows the profile for $\Omega_M=0.3$.  The dotted vertical lines show the location of splashback predicted by our toy model for $\Omega_M=1$ (blue) and $\Omega_M=0.3$ (red) respectively.   
\label{1dsim}}
\end{figure}

Our toy model also predicts a significant dependence on redshift (or equivalently, a dependence on the value of $\Omega_M$).  We cannot test that prediction using similarity solutions, because they assume $\Omega_M=1$.  To test this prediction, we therefore ran 1-dimensional N-body simulations of the collapse of isolated overdensities.  The simulations evolve the motion of spherical shells following Eqn. \eqref{EOM}.  The initial linear overdensity profiles are chosen to produce $M\propto a^s$ for various values of $s$.  Figure \ref{1dsim} shows an example, for $s=3$.  The solid curves in the figure show the results of the 1-D simulations, while for comparison, the dashed curve shows the similarity solution for $s=3$.  Note that for $\Omega_M=1$, the 1-D simulation does not exactly match the similarity solution.  This is because the dynamics, even in spherical symmetry, are subject to a slew of instabilities that are not present in the similarity solution \citep{Henriksen1997,Henriksen1999,Vogelsberger2011}.  To suppress these instabilities, we follow \citet{Vogelsberger2011} and soften the force law in Eqn.\ \eqref{EOM} near $r=0$.  As Fig.\ \ref{1dsim} shows, the halo profile for $\Omega_M=1$ is similar to the similarity solution.  The level of agreement or disagreement between the two curves illustrates the extent to which the 1-D N-body simulations may be trusted.  Note in particular that the location of the splashback radius is similar in the two cases.  The figure also shows results for $\Omega_M=0.3$, in the solid red curve.  For comparison, the vertical dotted lines show the toy model's predictions for the splashback radius for these values of $\Omega_M$.  Overall, we find good agreement, demonstrating that the location of splashback does indeed depend on cosmology and redshift.

Finally we compare with cosmological N-body simulations from the publicly available MultiDark Database.  The MDR1 simulation \citep{Multidark,Prada2012} contains $2048^3$ particles in a box of side length $L=1\, h^{-1} {\rm Gpc}$, giving a particle mass of $M_p=8.7\times 10^9 h^{-1} M_\odot$, while the Bolshoi simulation \citep{Klypin2011} contains $2048^3$ particles in a box of side length $L=250\, h^{-1} {\rm Mpc}$, giving a particle mass of $M_p=1.3\times 10^8 h^{-1} M_\odot$.  We extracted halos of mass $M_{\rm vir} \lesssim 10^{14} h^{-1} M_\odot$ from the Bolshoi simulation, and used MDR1 to obtain halos of mass $M\gtrsim 10^{14}  h^{-1} M_\odot$.  For both simulations, we used the publicly available Rockstar \citep{Rockstar} catalogs and merger trees \citep{Behroozi2013} \footnote{http://hipacc.ucsc.edu/Bolshoi/MergerTrees.html} to measure halo mass accretion histories.  For each halo, we walked the main branch of the merger tree to determine the mass accretion history (MAH) over a narrow redshift range, typically from $(0.67-1)\times a_{\rm obs}$. Then we fit the MAH to the form $M_{\rm vir} \propto e^{-\alpha z}$ \citep{Wechsler2002} over the narrow redshift range, and used the fitted value of $\alpha$ to determine $\Gamma=d(\log M_{\rm vir})/d(\log a)$ at $a=a_{\rm obs}$.  We then stacked halos in bins of mass, redshift, and $\Gamma$.

One point to note is that $\Gamma=d(\log M_{\rm vir})/d(\log a)$ need not be equal to the mass accretion rate $s=d(\log M)/d(\log a)$.  This is because $M_{\rm vir}$ does not represent the total amount of mass that has collapsed into a halo.  Instead, $M_{\rm vir}$ is the mass within a radius of average density $\Delta_{\rm vir}$.  As Fig.\ \ref{Delta_s} shows, the splashback radius (which does encompass the collapsed mass) occurs at various overdensity levels depending on redshift and accretion rate, meaning that sometimes $M_{\rm vir}$ will exceed $M_{\rm collapsed}$, sometimes $M_{\rm vir}$ will be less than $M_{\rm collapsed}$, and in general, the two will not evolve in the same way.  This difference in behavior has been termed pseudo-evolution by \citet{Diemer2013}, who point out that even in cases where there is no mass accretion and the halo mass profile is constant in physical units, the virial mass can still grow over time.  Because of pseudo-evolution, in general $\Gamma \neq s$.  This complicates the comparison of our model with cosmological N-body simulations.  Instead of stacking halos based on virial mass, we should stack halos based on the mass within their splashback radius, however it is difficult to determine the splashback radius for any individual halo.  We therefore stack the profiles of halos in bins of $M_{\rm vir}$ and $\Gamma$.  From the stacked profiles, we can determine the location of splashback and the average enclosed mass, $\langle M_{\rm collapsed}\rangle$.  By repeating this procedure for the progenitors of the stacked halos, we can determine $\langle M_{\rm collapsed}\rangle$ as a function of time, and thereby measure a typical accretion rate $s$ for each bin of $M_{\rm vir}$ and $\Gamma$.  We find that for $\Gamma \gtrsim 0.5$, the difference between the reconstructed $s$ and $\Gamma$ is small, typically of order $\Gamma-s \sim 0.1$.  We therefore use $\Gamma$ as a proxy for $s$, valid for $\Gamma>0.5$.

\begin{figure}
\centerline{\includegraphics[width=0.63\textwidth]{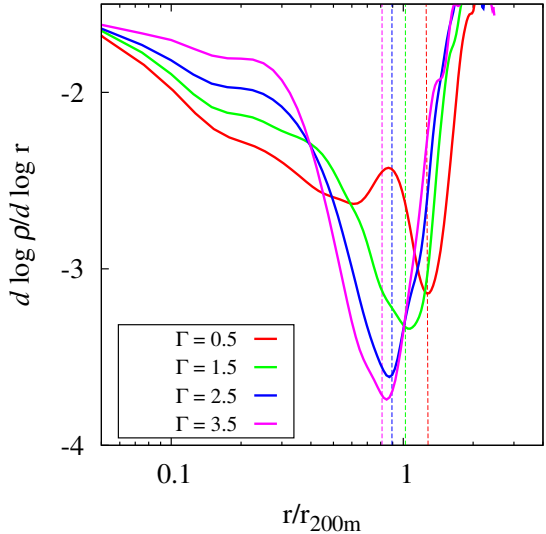}}
\caption{Dependence of splashback location on growth rate $\Gamma$.  Here we show the comparison between the model predictions and the simulation results for $z=0$ and $\Omega_M=0.27$.  The dashed vertical lines are at the predicted position of splashback as a function of $s\approx\Gamma$. The curves correspond to halos of mass $M_{\rm vir}=1-4\times10^{14} h^{-1} M_\odot$, binned according to growth rate $\Gamma$. \label{acc}}
\end{figure}

\begin{figure}
\centerline{\includegraphics[width=0.62\textwidth]{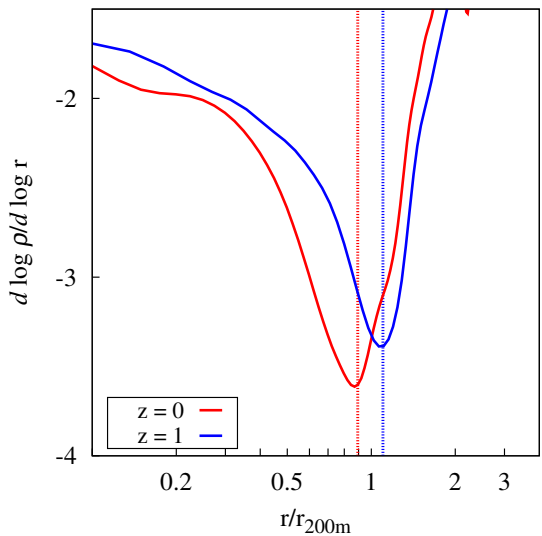}}
\caption{Redshift dependence of splashback location.  The blue curve shows $z=1$ halos of mass $M_{\rm vir} = 3 - 6\times10^{13} h^{-1} M_\odot$, while the red curve shows $z=0$ halos of mass $M_{\rm vir} = 1-4\times10^{14}  h^{-1}M_\odot$.  Both samples have accretion rates $\Gamma=2.5$.  The vertical lines indicate the expected splashback radii.\label{z}}
\end{figure}

\begin{figure*}
\includegraphics[width=0.9\textwidth, trim= 0in 0.5in 0in 0.5in,clip]{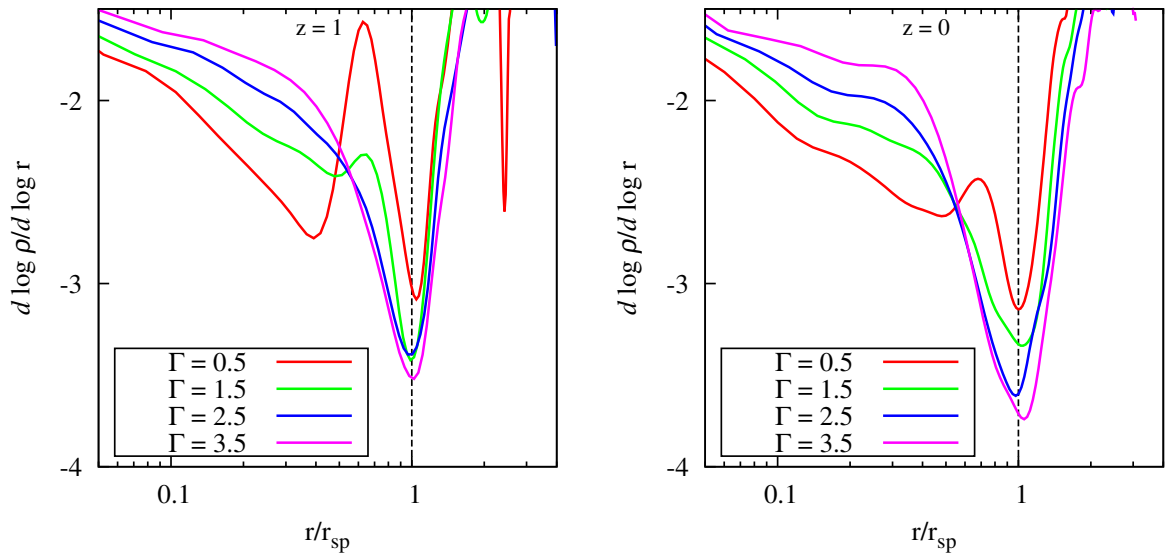}
\caption{(Left) Local density slope as a function of $r/r_{\rm sp}$ for stacked N-body halos at $z=1$ with mass $M_{\rm vir}=3- 6\times 10^{13}h^{-1} M_\odot$, where $r_{\rm sp}$ is the predicted splashback radius computed using the spherical collapse model of \S\ref{toy}.  Different colors correspond to different accretion rates as indicated.  The vertical black dashed line shows the expected position of splashback.  (Right) Similar, but for halos with $M_{\rm vir}=1- 4\times 10^{14}h^{-1} M_\odot$ at $z=0$. \label{s}}
\end{figure*}

\begin{figure}[t]
\centerline{\includegraphics[width=0.59\textwidth]{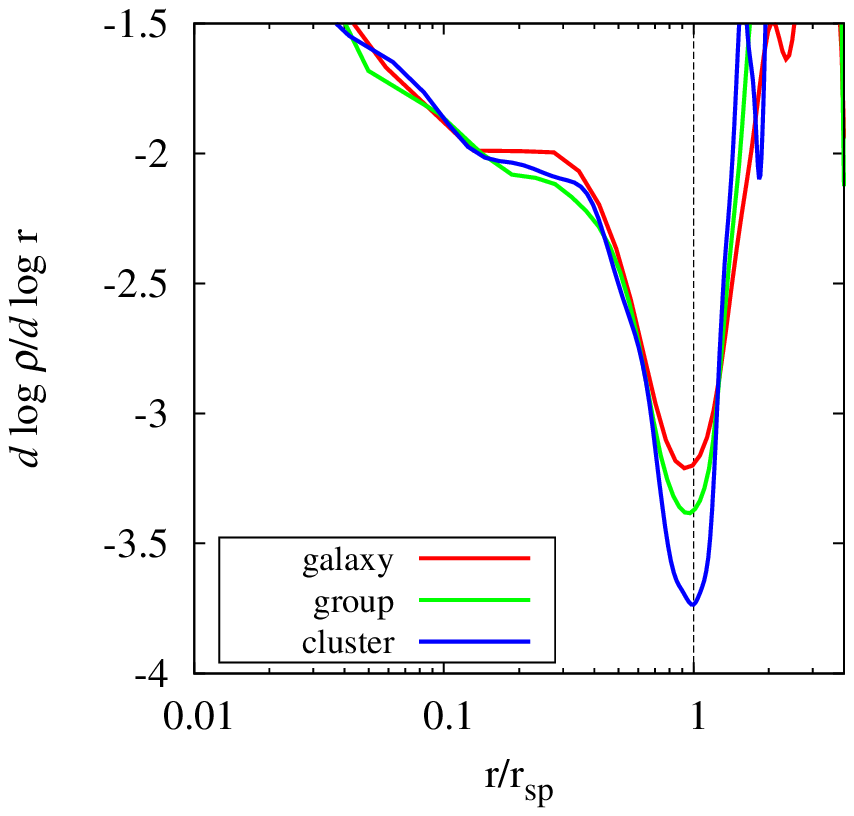}}
\caption{Local density slope for halos with accretion rate $\Gamma=2$ and $z=0$, as a function of halo mass or $\nu$.  The various curves correspond to $M_{\rm vir}= 1-3 \times 10^{12} h^{-1} M_\odot $ (red),  $M_{\rm vir}= 3 - 5 \times 10^{13} h^{-1} M_\odot$ (green), and $M_{\rm vir}=3-4 \times 10^{14} h^{-1} M_\odot$ (blue).  These mass bins correspond to $\nu\sim 0.9$, 1.5, and 2.5, respectively.  The location of the splashback feature remains the same, independent of mass, although the depth of the feature depends on $M$ (or equivalently $\nu$). 
\label{m}}
\end{figure}

With that caveat in mind, we now move on to the comparison of our model with cosmological N-body simulations.  Our results are shown in Figs.\ \ref{acc}-\ref{m}.  Figure \ref{acc} shows the dependence of the splashback radius on the accretion rate, $\Gamma$.  As shown by \citet{Diemer2014}, splashback occurs at a higher density (smaller $r/r_{200m}$) for higher accretion rates.   Our model's prediction for $r_{\rm sp}$ agrees quite well with the observed dependence on accretion rate, across the entire range we have checked ($0.5 < \Gamma < 3.5$).  We also confirm the model's predicted redshift dependence of splashback (Fig.\ \ref{z}).  Since these figures have many overlapping curves at different radii, it may be difficult to see how well the observed splashback radius agrees with the model prediction.  Therefore, in Fig.\ \ref{s} we scale $r$ by the predicted splashback radius $r_{\rm sp}$, and show that splashback occurs where predicted, at all accretion rates and redshifts.  Similarly, as predicted by the model, the location of splashback does not depend on halo mass $M$ (or equivalently, $\nu\equiv \delta_c/\sigma(M)$), as illustrated in figure \ref{m}.

One possibly interesting feature shown in Fig.\ \ref{m}, and also noted by \citet{Diemer2014}, is that although the location of the steepening feature is independent of $M$ or $\nu$, the {\em depth} of the feature does depend systematically on $\nu$.  As shown in the figure, the lower mass halos, which correspond to peaks of smaller $\nu$, have systematically shallower caustics.  The behavior is reminiscent of that shown in Fig.\ \ref{ssim}, which showed that similarity solutions with stronger initial triaxiality produce caustics that are progressively more non-spherical, producing shallower features in the spherically averaged profile.  Similar behavior could also explain the $\nu$ dependence of the N-body halos.  It is well known from the statistics of peaks of Gaussian random fields that peaks of larger height $\nu$ are systematically more spherical, with a smaller range of ellipticity $e$, than peaks of low $\nu$ \citep{BBKS}.  Based on the behavior shown in the similarity solutions shown in Fig. \ref{ssim}, we might therefore expect that N-body halos of larger $\nu$ will have smaller initial ellipticity, and therefore have deeper caustics, than halos with lower $\nu$, exactly as found in Fig.\ \ref{m}.  We have not investigated this topic in this paper, but it may be worth exploring in future work.  

\begin{figure*}
\centerline{
\includegraphics[width=0.33\textwidth, trim=1.7in 1.5cm 1.7in 0.5cm]{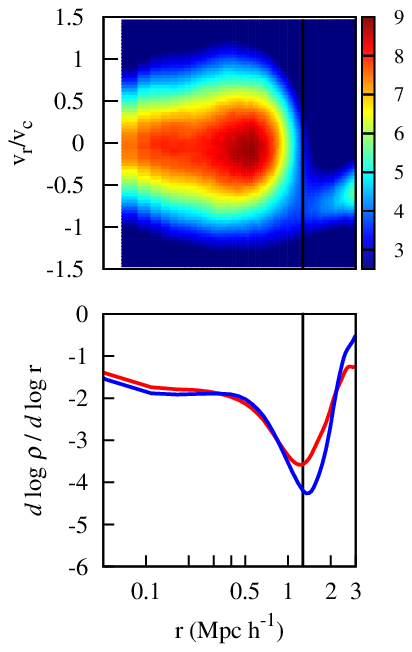} 
\includegraphics[width=0.33\textwidth, trim=1.7in 1.5cm 1.7in 0.5cm]{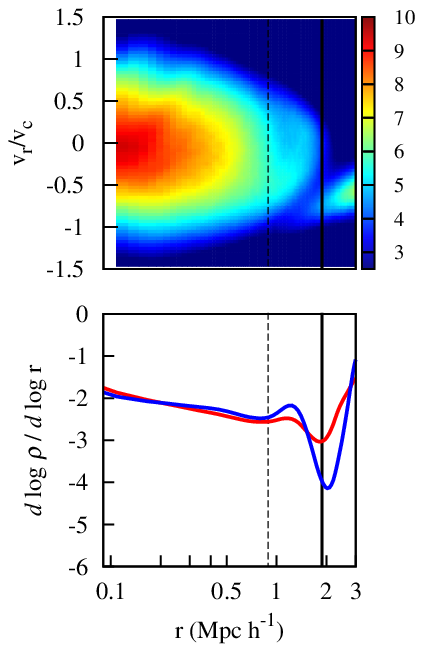} 
\includegraphics[width=0.33\textwidth, trim=1.7in 1.5cm 1.7in 0.5cm]{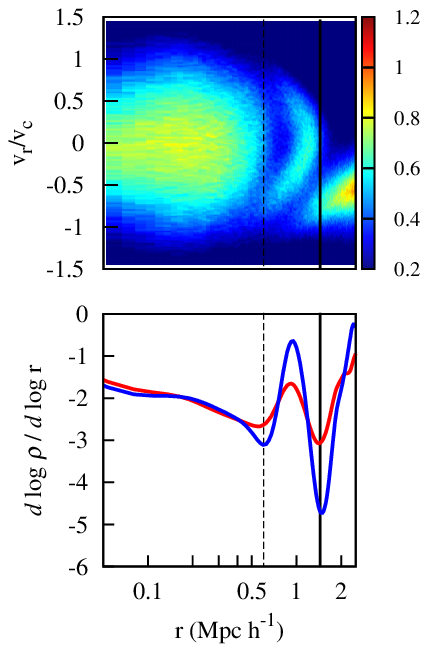} 
}
\caption{Phase space diagram for different accretion rates.  The left panel shows halos in the mass range $M_{\rm vir}=1-4\times 10^{14} h^{-1} M_\odot$ at $z=0$ for accretion rates $\Gamma\approx 3$.  The middle panel shows halos of the same mass and redshift, but with $\Gamma\approx 0.5$.  The right panel shows halos with $M_{\rm vir}=3-6\times 10^{13} h^{-1} M_\odot$ at $z=1$, also with $\Gamma\approx 0.5$.  The bottom panels show the slope of the local density profile of all mass in red, and for particles with $|v_r|<0.4 v_c$ in blue.  This is similar to Fig.\ \ref{phasespace}, which showed results for an intermediate growth rate $\Gamma \approx 1.5$.  Note that at low accretion rates, the splashback material forms a distinct stream, which leads to multiple minima in the run of density slope vs.\ radius, indicated by the vertical dashed lines.
\label{2caustic}
}
\end{figure*}

At low accretion rates, $\Gamma < 1$, the stacked profiles begin to exhibit additional features besides the steepening at $r_{\rm sp}$.  The most pronounced example of this is the stacked profile for $\Gamma\approx 0.5$, shown in Fig.\ \ref{s}.  Instead of monotonically becoming more shallow at $r<r_{\rm sp}$, the slope of the density profile oscillates.  The origin of this behavior may be understood by examining the ensemble phase space diagram for these halos, shown in Figure \ref{2caustic}.  The phase space structure for low accretion rate is distinct from the other $\Gamma$ bins, in that the stream of splashback material is noticeably separated from the rest of the virialized matter in the halo.  This behavior is similar to the phase space structure seen in spherical self-similar collapse \citep{FG84}, in which each separate stream produces a separate caustic (see Fig.\ \ref{ssim}).

\section{Discussion}

\begin{figure}
\centerline{\includegraphics[width=0.7\textwidth, trim=0in 0.2in 0in 0.5in]{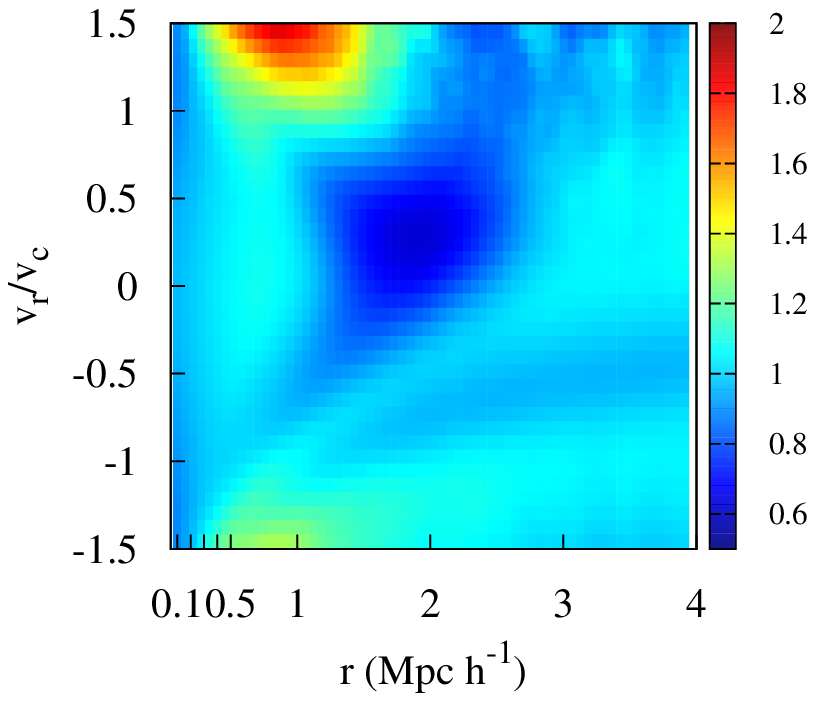}}
\caption{Map of accretion rate $\Gamma$ as a function of phase space coordinates $r$ and $v_r$ for halos in the mass range $M_{\rm vir}=1-2\times 10^{14} h^{-1} M_\odot$ at $z=0$.  For these halos, $R_{\rm vir}\approx 1 h^{-1} {\rm Mpc}$.  Note that at large radii, $r\sim 2 h^{-1} {\rm Mpc}$, most of the outgoing material ($v_r>0$) is associated with slowly accreting halos with $\Gamma \approx 0.5$.  
\label{smap}}
\end{figure}

We have presented an extremely simple model, derived from first principles, that explains the location of splashback around cosmological N-body halos.  The model is a simple extension to the standard spherical collapse model \citep{GunnGott72}.  Although the model has no free parameters, it nevertheless accurately predicts the location of splashback for halos of a variety of different masses, redshifts, and accretion rates.  The splashback radius depends principally on the mass accretion rate $s=d\log M/d\log a$, as found by \citep{Diemer2014}.  However, it also depends on redshift, due to the variation in the value of $\Omega_M$ over time.

Depending on the accretion rate and redshift, the overdensity at the splashback radius, $\Delta_s$, can be much larger or much smaller than either $\Delta=200$ or $\Delta_{\rm vir}$, the nominal `virial' overdensity, which is typically in the range 200-300 \citep{Eke1996}.  Since the splashback radius encompasses the multi-streaming region, this means that the virialized region surrounding a halo can extend far beyond $R_{\rm vir}$.  This fact is not surprising: numerous simulations have found that material which has passed inside $R_{\rm vir}$ can later be found at large distance, e.g.\ $\sim 2-3 R_{\rm vir}$ \citep{Gill2005,Diemand2008,Ludlow2009,Anderhalden2011,Wetzel2014,Garrison-Kimmel2014}.  As we have seen, recently accreted material can splash back to such large distances, when the accretion rate in the halo is low.  Our expectation, therefore, would be that much of the halo material found at large distances ($\sim 2-3 R_{\rm vir}$) is associated with slowly accreting halos.  To demonstrate this, we plot in Figure \ref{smap} the average halo accretion rate for material around halos of mass $M_{\rm vir}=1-4\times 10^{14} h^{-1} M_\odot$ at $z=0$ as a function of phase space coordinates $r$ and $v_r$.  Outgoing material (i.e.\ $v_r>0$) at $r\approx 2R_{\rm vir}$ (which is $r\sim 2 h^{-1} {\rm Mpc}$ for this mass bin) is typically found around halos with the lowest accretion rates, $\Gamma \approx 0.5$. 

In principle, the steepening associated with splashback could be observable.  The projected surface density of dark matter halos shows a similar steepening at the same location at the 3D steepening feature \citep{Diemer2014}, implying that this feature may be observable in the stacked lensing profiles of ensembles of halos \citep{Oguri2011}.  As we have seen, the steepening feature associated with splashback is most prominent for massive (high $\nu$) objects with high accretion rates.  This suggests focusing on galaxy clusters, rather than lower-mass objects like galaxies whose lower accretion rates would produce weak splashback features possibly in the 2-halo region.  In practice, it is often difficult to predict halo masses from observable quantities like cluster richness or SZ decrement, meaning that stacked profiles will typically average over wide mass bins.  This will not necessarily wash out the steepening signature, however.  In our analysis of stacked profiles, we found that stacking rather wide bins of mass (such as a factor of 4 in $M_{\rm vir}$) does not wash out the steepening feature, e.g.\ Fig.\ \ref{s}.  Therefore, realistic uncertainties in the mass-observable relation should not wash out the steepening feature entirely.  If the steepening feature is measured with high significance, then it may be interesting to stack halos as a function of various observable properties like concentration.  Measuring how the splashback radius depends on those observable properties immediately translates into a measurement of how well the mass accretion rate correlates with those properties, which can test the predicted behavior for CDM cosmologies \citep{Wechsler2002,Dalal08}.  

Similar steepening features could also arise in the baryonic components of halos.  Unlike the dark matter, however, the baryons are not collisionless, which means that splashback need not occur at the same location as the dark matter.  Hydrodynamic simulations of galaxy cluster outskirts can quantify whether gas splashback occurs near dark matter splashback \citep{Lau2014}.  Stars, unlike gas, are effectively collisionless at the densities found in halo outskirts, so the stacked starlight profiles of ensembles of halos could in principle exhibit similar behavior as the dark matter.  Dynamical friction could potentially slow down stars in galaxies relative to unbound dark matter, though, so it may be worthwhile performing simulations with star formation to check if stars produce similar caustics as the dark matter.

\begin{acknowledgments}
We thank Andrey Kravtsov and Benedikt Diemer for many helpful discussions and for a careful reading of an earlier version of this paper.  
ND is supported by NASA under grants NNX12AD02G and NNX12AC99G, and by a Sloan Fellowship.
The MultiDark Database used in this paper and the web application providing online access to it
were constructed as part of the activities of the German Astrophysical Virtual Observatory as result
of a collaboration between the Leibniz-Institute for Astrophysics Potsdam (AIP) and the Spanish
MultiDark Consolider Project CSD2009-00064. The Bolshoi and MultiDark simulations were run on 
NASA's Pleiades supercomputer at the NASA Ames Research Center.
\end{acknowledgments}

\newcommand{\aj}{Astron.\ J.}
\newcommand{\apjl}{\apj\ Lett.}
\newcommand{\jcap}{Journal of Cosmology and Astroparticle Physics}
\newcommand{\mnras}{MNRAS}

%\bibliographystyle{apsrev4-1}
%\bibliography{splash}

\begin{thebibliography}{37}%
\makeatletter
\providecommand \@ifxundefined [1]{%
 \@ifx{#1\undefined}
}%
\providecommand \@ifnum [1]{%
 \ifnum #1\expandafter \@firstoftwo
 \else \expandafter \@secondoftwo
 \fi
}%
\providecommand \@ifx [1]{%
 \ifx #1\expandafter \@firstoftwo
 \else \expandafter \@secondoftwo
 \fi
}%
\providecommand \natexlab [1]{#1}%
\providecommand \enquote  [1]{``#1''}%
\providecommand \bibnamefont  [1]{#1}%
\providecommand \bibfnamefont [1]{#1}%
\providecommand \citenamefont [1]{#1}%
\providecommand \href@noop [0]{\@secondoftwo}%
\providecommand \href [0]{\begingroup \@sanitize@url \@href}%
\providecommand \@href[1]{\@@startlink{#1}\@@href}%
\providecommand \@@href[1]{\endgroup#1\@@endlink}%
\providecommand \@sanitize@url [0]{\catcode `\\12\catcode `\$12\catcode
  `\&12\catcode `\#12\catcode `\^12\catcode `\_12\catcode `\%12\relax}%
\providecommand \@@startlink[1]{}%
\providecommand \@@endlink[0]{}%
\providecommand \url  [0]{\begingroup\@sanitize@url \@url }%
\providecommand \@url [1]{\endgroup\@href {#1}{\urlprefix }}%
\providecommand \urlprefix  [0]{URL }%
\providecommand \Eprint [0]{\href }%
\providecommand \doibase [0]{http://dx.doi.org/}%
\providecommand \selectlanguage [0]{\@gobble}%
\providecommand \bibinfo  [0]{\@secondoftwo}%
\providecommand \bibfield  [0]{\@secondoftwo}%
\providecommand \translation [1]{[#1]}%
\providecommand \BibitemOpen [0]{}%
\providecommand \bibitemStop [0]{}%
\providecommand \bibitemNoStop [0]{.\EOS\space}%
\providecommand \EOS [0]{\spacefactor3000\relax}%
\providecommand \BibitemShut  [1]{\csname bibitem#1\endcsname}%
\let\auto@bib@innerbib\@empty
%</preamble>
\bibitem [{\citenamefont {{Gunn}}\ and\ \citenamefont
  {{Gott}}(1972)}]{GunnGott72}%
  \BibitemOpen
  \bibfield  {author} {\bibinfo {author} {\bibfnamefont {J.~E.}\ \bibnamefont
  {{Gunn}}}\ and\ \bibinfo {author} {\bibfnamefont {J.~R.~I.}\ \bibnamefont
  {{Gott}}},\ }\href {\doibase 10.1086/151605} {\bibfield  {journal} {\bibinfo
  {journal} {\apj}\ }\textbf {\bibinfo {volume} {176}},\ \bibinfo {pages} {1}
  (\bibinfo {year} {1972})}\BibitemShut {NoStop}%
\bibitem [{\citenamefont {{Fillmore}}\ and\ \citenamefont
  {{Goldreich}}(1984)}]{FG84}%
  \BibitemOpen
  \bibfield  {author} {\bibinfo {author} {\bibfnamefont {J.~A.}\ \bibnamefont
  {{Fillmore}}}\ and\ \bibinfo {author} {\bibfnamefont {P.}~\bibnamefont
  {{Goldreich}}},\ }\href {\doibase 10.1086/162070} {\bibfield  {journal}
  {\bibinfo  {journal} {\apj}\ }\textbf {\bibinfo {volume} {281}},\ \bibinfo
  {pages} {1} (\bibinfo {year} {1984})}\BibitemShut {NoStop}%
\bibitem [{\citenamefont {{Diemand}}\ \emph {et~al.}(2007)\citenamefont
  {{Diemand}}, \citenamefont {{Kuhlen}},\ and\ \citenamefont
  {{Madau}}}]{ViaLactea1}%
  \BibitemOpen
  \bibfield  {author} {\bibinfo {author} {\bibfnamefont {J.}~\bibnamefont
  {{Diemand}}}, \bibinfo {author} {\bibfnamefont {M.}~\bibnamefont {{Kuhlen}}},
  \ and\ \bibinfo {author} {\bibfnamefont {P.}~\bibnamefont {{Madau}}},\ }\href
  {\doibase 10.1086/520573} {\bibfield  {journal} {\bibinfo  {journal} {\apj}\
  }\textbf {\bibinfo {volume} {667}},\ \bibinfo {pages} {859} (\bibinfo {year}
  {2007})},\ \Eprint {http://arxiv.org/abs/astro-ph/0703337} {astro-ph/0703337}
  \BibitemShut {NoStop}%
\bibitem [{\citenamefont {{Diemand}}\ \emph {et~al.}(2008)\citenamefont
  {{Diemand}}, \citenamefont {{Kuhlen}}, \citenamefont {{Madau}}, \citenamefont
  {{Zemp}}, \citenamefont {{Moore}}, \citenamefont {{Potter}},\ and\
  \citenamefont {{Stadel}}}]{ViaLactea2}%
  \BibitemOpen
  \bibfield  {author} {\bibinfo {author} {\bibfnamefont {J.}~\bibnamefont
  {{Diemand}}}, \bibinfo {author} {\bibfnamefont {M.}~\bibnamefont {{Kuhlen}}},
  \bibinfo {author} {\bibfnamefont {P.}~\bibnamefont {{Madau}}}, \bibinfo
  {author} {\bibfnamefont {M.}~\bibnamefont {{Zemp}}}, \bibinfo {author}
  {\bibfnamefont {B.}~\bibnamefont {{Moore}}}, \bibinfo {author} {\bibfnamefont
  {D.}~\bibnamefont {{Potter}}}, \ and\ \bibinfo {author} {\bibfnamefont
  {J.}~\bibnamefont {{Stadel}}},\ }\href {\doibase 10.1038/nature07153}
  {\bibfield  {journal} {\bibinfo  {journal} {\nat}\ }\textbf {\bibinfo
  {volume} {454}},\ \bibinfo {pages} {735} (\bibinfo {year} {2008})},\ \Eprint
  {http://arxiv.org/abs/0805.1244} {arXiv:0805.1244} \BibitemShut {NoStop}%
\bibitem [{\citenamefont {{Stadel}}\ \emph {et~al.}(2009)\citenamefont
  {{Stadel}}, \citenamefont {{Potter}}, \citenamefont {{Moore}}, \citenamefont
  {{Diemand}}, \citenamefont {{Madau}}, \citenamefont {{Zemp}}, \citenamefont
  {{Kuhlen}},\ and\ \citenamefont {{Quilis}}}]{GHalo}%
  \BibitemOpen
  \bibfield  {author} {\bibinfo {author} {\bibfnamefont {J.}~\bibnamefont
  {{Stadel}}}, \bibinfo {author} {\bibfnamefont {D.}~\bibnamefont {{Potter}}},
  \bibinfo {author} {\bibfnamefont {B.}~\bibnamefont {{Moore}}}, \bibinfo
  {author} {\bibfnamefont {J.}~\bibnamefont {{Diemand}}}, \bibinfo {author}
  {\bibfnamefont {P.}~\bibnamefont {{Madau}}}, \bibinfo {author} {\bibfnamefont
  {M.}~\bibnamefont {{Zemp}}}, \bibinfo {author} {\bibfnamefont
  {M.}~\bibnamefont {{Kuhlen}}}, \ and\ \bibinfo {author} {\bibfnamefont
  {V.}~\bibnamefont {{Quilis}}},\ }\href {\doibase
  10.1111/j.1745-3933.2009.00699.x} {\bibfield  {journal} {\bibinfo  {journal}
  {\mnras}\ }\textbf {\bibinfo {volume} {398}},\ \bibinfo {pages} {L21}
  (\bibinfo {year} {2009})},\ \Eprint {http://arxiv.org/abs/0808.2981}
  {arXiv:0808.2981} \BibitemShut {NoStop}%
\bibitem [{\citenamefont {{Navarro}}\ \emph {et~al.}(2010)\citenamefont
  {{Navarro}}, \citenamefont {{Ludlow}}, \citenamefont {{Springel}},
  \citenamefont {{Wang}}, \citenamefont {{Vogelsberger}}, \citenamefont
  {{White}}, \citenamefont {{Jenkins}}, \citenamefont {{Frenk}},\ and\
  \citenamefont {{Helmi}}}]{Aquarius}%
  \BibitemOpen
  \bibfield  {author} {\bibinfo {author} {\bibfnamefont {J.~F.}\ \bibnamefont
  {{Navarro}}}, \bibinfo {author} {\bibfnamefont {A.}~\bibnamefont {{Ludlow}}},
  \bibinfo {author} {\bibfnamefont {V.}~\bibnamefont {{Springel}}}, \bibinfo
  {author} {\bibfnamefont {J.}~\bibnamefont {{Wang}}}, \bibinfo {author}
  {\bibfnamefont {M.}~\bibnamefont {{Vogelsberger}}}, \bibinfo {author}
  {\bibfnamefont {S.~D.~M.}\ \bibnamefont {{White}}}, \bibinfo {author}
  {\bibfnamefont {A.}~\bibnamefont {{Jenkins}}}, \bibinfo {author}
  {\bibfnamefont {C.~S.}\ \bibnamefont {{Frenk}}}, \ and\ \bibinfo {author}
  {\bibfnamefont {A.}~\bibnamefont {{Helmi}}},\ }\href {\doibase
  10.1111/j.1365-2966.2009.15878.x} {\bibfield  {journal} {\bibinfo  {journal}
  {\mnras}\ }\textbf {\bibinfo {volume} {402}},\ \bibinfo {pages} {21}
  (\bibinfo {year} {2010})},\ \Eprint {http://arxiv.org/abs/0810.1522}
  {arXiv:0810.1522} \BibitemShut {NoStop}%
\bibitem [{\citenamefont {{Lithwick}}\ and\ \citenamefont
  {{Dalal}}(2011)}]{LD10}%
  \BibitemOpen
  \bibfield  {author} {\bibinfo {author} {\bibfnamefont {Y.}~\bibnamefont
  {{Lithwick}}}\ and\ \bibinfo {author} {\bibfnamefont {N.}~\bibnamefont
  {{Dalal}}},\ }\href {\doibase 10.1088/0004-637X/734/2/100} {\bibfield
  {journal} {\bibinfo  {journal} {\apj}\ }\textbf {\bibinfo {volume} {734}},\
  \bibinfo {eid} {100} (\bibinfo {year} {2011})},\ \Eprint
  {http://arxiv.org/abs/1010.3723} {arXiv:1010.3723 [astro-ph.CO]} \BibitemShut
  {NoStop}%
\bibitem [{\citenamefont {{Dalal}}\ \emph {et~al.}(2010)\citenamefont
  {{Dalal}}, \citenamefont {{Lithwick}},\ and\ \citenamefont
  {{Kuhlen}}}]{Dalal2010}%
  \BibitemOpen
  \bibfield  {author} {\bibinfo {author} {\bibfnamefont {N.}~\bibnamefont
  {{Dalal}}}, \bibinfo {author} {\bibfnamefont {Y.}~\bibnamefont {{Lithwick}}},
  \ and\ \bibinfo {author} {\bibfnamefont {M.}~\bibnamefont {{Kuhlen}}},\
  }\href@noop {} {\bibfield  {journal} {\bibinfo  {journal} {ArXiv e-prints}\ }
  (\bibinfo {year} {2010})},\ \Eprint {http://arxiv.org/abs/1010.2539}
  {arXiv:1010.2539 [astro-ph.CO]} \BibitemShut {NoStop}%
\bibitem [{\citenamefont {{Diemer}}\ and\ \citenamefont
  {{Kravtsov}}(2014)}]{Diemer2014}%
  \BibitemOpen
  \bibfield  {author} {\bibinfo {author} {\bibfnamefont {B.}~\bibnamefont
  {{Diemer}}}\ and\ \bibinfo {author} {\bibfnamefont {A.~V.}\ \bibnamefont
  {{Kravtsov}}},\ }\href {\doibase 10.1088/0004-637X/789/1/1} {\bibfield
  {journal} {\bibinfo  {journal} {\apj}\ }\textbf {\bibinfo {volume} {789}},\
  \bibinfo {eid} {1} (\bibinfo {year} {2014})},\ \Eprint
  {http://arxiv.org/abs/1401.1216} {arXiv:1401.1216} \BibitemShut {NoStop}%
\bibitem [{\citenamefont {{Navarro}}\ \emph {et~al.}(1996)\citenamefont
  {{Navarro}}, \citenamefont {{Frenk}},\ and\ \citenamefont {{White}}}]{NFW96}%
  \BibitemOpen
  \bibfield  {author} {\bibinfo {author} {\bibfnamefont {J.~F.}\ \bibnamefont
  {{Navarro}}}, \bibinfo {author} {\bibfnamefont {C.~S.}\ \bibnamefont
  {{Frenk}}}, \ and\ \bibinfo {author} {\bibfnamefont {S.~D.~M.}\ \bibnamefont
  {{White}}},\ }\href {\doibase 10.1086/177173} {\bibfield  {journal} {\bibinfo
   {journal} {\apj}\ }\textbf {\bibinfo {volume} {462}},\ \bibinfo {pages}
  {563} (\bibinfo {year} {1996})},\ \Eprint
  {http://arxiv.org/abs/astro-ph/9508025} {astro-ph/9508025} \BibitemShut
  {NoStop}%
\bibitem [{\citenamefont {{Navarro}}\ \emph {et~al.}(1997)\citenamefont
  {{Navarro}}, \citenamefont {{Frenk}},\ and\ \citenamefont {{White}}}]{NFW97}%
  \BibitemOpen
  \bibfield  {author} {\bibinfo {author} {\bibfnamefont {J.~F.}\ \bibnamefont
  {{Navarro}}}, \bibinfo {author} {\bibfnamefont {C.~S.}\ \bibnamefont
  {{Frenk}}}, \ and\ \bibinfo {author} {\bibfnamefont {S.~D.~M.}\ \bibnamefont
  {{White}}},\ }\href@noop {} {\bibfield  {journal} {\bibinfo  {journal}
  {\apj}\ }\textbf {\bibinfo {volume} {490}},\ \bibinfo {pages} {493} (\bibinfo
  {year} {1997})},\ \Eprint {http://arxiv.org/abs/astro-ph/9611107}
  {astro-ph/9611107} \BibitemShut {NoStop}%
\bibitem [{\citenamefont {{Merritt}}\ \emph {et~al.}(2005)\citenamefont
  {{Merritt}}, \citenamefont {{Navarro}}, \citenamefont {{Ludlow}},\ and\
  \citenamefont {{Jenkins}}}]{Merritt05}%
  \BibitemOpen
  \bibfield  {author} {\bibinfo {author} {\bibfnamefont {D.}~\bibnamefont
  {{Merritt}}}, \bibinfo {author} {\bibfnamefont {J.~F.}\ \bibnamefont
  {{Navarro}}}, \bibinfo {author} {\bibfnamefont {A.}~\bibnamefont {{Ludlow}}},
  \ and\ \bibinfo {author} {\bibfnamefont {A.}~\bibnamefont {{Jenkins}}},\
  }\href {\doibase 10.1086/430636} {\bibfield  {journal} {\bibinfo  {journal}
  {\apjl}\ }\textbf {\bibinfo {volume} {624}},\ \bibinfo {pages} {L85}
  (\bibinfo {year} {2005})},\ \Eprint {http://arxiv.org/abs/astro-ph/0502515}
  {astro-ph/0502515} \BibitemShut {NoStop}%
\bibitem [{\citenamefont {{Merritt}}\ \emph {et~al.}(2006)\citenamefont
  {{Merritt}}, \citenamefont {{Graham}}, \citenamefont {{Moore}}, \citenamefont
  {{Diemand}},\ and\ \citenamefont {{Terzi{\'c}}}}]{Merritt06}%
  \BibitemOpen
  \bibfield  {author} {\bibinfo {author} {\bibfnamefont {D.}~\bibnamefont
  {{Merritt}}}, \bibinfo {author} {\bibfnamefont {A.~W.}\ \bibnamefont
  {{Graham}}}, \bibinfo {author} {\bibfnamefont {B.}~\bibnamefont {{Moore}}},
  \bibinfo {author} {\bibfnamefont {J.}~\bibnamefont {{Diemand}}}, \ and\
  \bibinfo {author} {\bibfnamefont {B.}~\bibnamefont {{Terzi{\'c}}}},\ }\href
  {\doibase 10.1086/508988} {\bibfield  {journal} {\bibinfo  {journal} {\aj}\
  }\textbf {\bibinfo {volume} {132}},\ \bibinfo {pages} {2685} (\bibinfo {year}
  {2006})},\ \Eprint {http://arxiv.org/abs/astro-ph/0509417} {astro-ph/0509417}
  \BibitemShut {NoStop}%
\bibitem [{\citenamefont {{Diemand}}\ and\ \citenamefont
  {{Kuhlen}}(2008)}]{Diemand2008}%
  \BibitemOpen
  \bibfield  {author} {\bibinfo {author} {\bibfnamefont {J.}~\bibnamefont
  {{Diemand}}}\ and\ \bibinfo {author} {\bibfnamefont {M.}~\bibnamefont
  {{Kuhlen}}},\ }\href {\doibase 10.1086/589688} {\bibfield  {journal}
  {\bibinfo  {journal} {\apjl}\ }\textbf {\bibinfo {volume} {680}},\ \bibinfo
  {pages} {L25} (\bibinfo {year} {2008})},\ \Eprint
  {http://arxiv.org/abs/0804.4185} {arXiv:0804.4185} \BibitemShut {NoStop}%
\bibitem [{\citenamefont {{Vogelsberger}}\ \emph {et~al.}(2011)\citenamefont
  {{Vogelsberger}}, \citenamefont {{Mohayaee}},\ and\ \citenamefont
  {{White}}}]{Vogelsberger2011}%
  \BibitemOpen
  \bibfield  {author} {\bibinfo {author} {\bibfnamefont {M.}~\bibnamefont
  {{Vogelsberger}}}, \bibinfo {author} {\bibfnamefont {R.}~\bibnamefont
  {{Mohayaee}}}, \ and\ \bibinfo {author} {\bibfnamefont {S.~D.~M.}\
  \bibnamefont {{White}}},\ }\href {\doibase 10.1111/j.1365-2966.2011.18605.x}
  {\bibfield  {journal} {\bibinfo  {journal} {\mnras}\ }\textbf {\bibinfo
  {volume} {414}},\ \bibinfo {pages} {3044} (\bibinfo {year} {2011})},\ \Eprint
  {http://arxiv.org/abs/1007.4195} {arXiv:1007.4195 [astro-ph.CO]} \BibitemShut
  {NoStop}%
\bibitem [{\citenamefont {{Cooper}}\ \emph {et~al.}(2011)\citenamefont
  {{Cooper}}, \citenamefont {{Mart{\'{\i}}nez-Delgado}}, \citenamefont
  {{Helly}}, \citenamefont {{Frenk}}, \citenamefont {{Cole}}, \citenamefont
  {{Crawford}}, \citenamefont {{Zibetti}}, \citenamefont {{Carballo-Bello}},\
  and\ \citenamefont {{GaBany}}}]{Cooper2011}%
  \BibitemOpen
  \bibfield  {author} {\bibinfo {author} {\bibfnamefont {A.~P.}\ \bibnamefont
  {{Cooper}}}, \bibinfo {author} {\bibfnamefont {D.}~\bibnamefont
  {{Mart{\'{\i}}nez-Delgado}}}, \bibinfo {author} {\bibfnamefont
  {J.}~\bibnamefont {{Helly}}}, \bibinfo {author} {\bibfnamefont
  {C.}~\bibnamefont {{Frenk}}}, \bibinfo {author} {\bibfnamefont
  {S.}~\bibnamefont {{Cole}}}, \bibinfo {author} {\bibfnamefont
  {K.}~\bibnamefont {{Crawford}}}, \bibinfo {author} {\bibfnamefont
  {S.}~\bibnamefont {{Zibetti}}}, \bibinfo {author} {\bibfnamefont {J.~A.}\
  \bibnamefont {{Carballo-Bello}}}, \ and\ \bibinfo {author} {\bibfnamefont
  {R.~J.}\ \bibnamefont {{GaBany}}},\ }\href {\doibase
  10.1088/2041-8205/743/1/L21} {\bibfield  {journal} {\bibinfo  {journal}
  {\apjl}\ }\textbf {\bibinfo {volume} {743}},\ \bibinfo {eid} {L21} (\bibinfo
  {year} {2011})},\ \Eprint {http://arxiv.org/abs/1111.2864} {arXiv:1111.2864
  [astro-ph.GA]} \BibitemShut {NoStop}%
\bibitem [{Note1()}]{Note1}%
  \BibitemOpen
  \bibinfo {note} {http://www.multidark.org/MultiDark}\BibitemShut {NoStop}%
\bibitem [{\citenamefont {{Eke}}\ \emph {et~al.}(1996)\citenamefont {{Eke}},
  \citenamefont {{Cole}},\ and\ \citenamefont {{Frenk}}}]{Eke1996}%
  \BibitemOpen
  \bibfield  {author} {\bibinfo {author} {\bibfnamefont {V.~R.}\ \bibnamefont
  {{Eke}}}, \bibinfo {author} {\bibfnamefont {S.}~\bibnamefont {{Cole}}}, \
  and\ \bibinfo {author} {\bibfnamefont {C.~S.}\ \bibnamefont {{Frenk}}},\
  }\href@noop {} {\bibfield  {journal} {\bibinfo  {journal} {\mnras}\ }\textbf
  {\bibinfo {volume} {282}},\ \bibinfo {pages} {263} (\bibinfo {year}
  {1996})},\ \Eprint {http://arxiv.org/abs/astro-ph/9601088} {astro-ph/9601088}
  \BibitemShut {NoStop}%
\bibitem [{\citenamefont {{Henriksen}}\ and\ \citenamefont
  {{Widrow}}(1997)}]{Henriksen1997}%
  \BibitemOpen
  \bibfield  {author} {\bibinfo {author} {\bibfnamefont {R.~N.}\ \bibnamefont
  {{Henriksen}}}\ and\ \bibinfo {author} {\bibfnamefont {L.~M.}\ \bibnamefont
  {{Widrow}}},\ }\href {\doibase 10.1103/PhysRevLett.78.3426} {\bibfield
  {journal} {\bibinfo  {journal} {Physical Review Letters}\ }\textbf {\bibinfo
  {volume} {78}},\ \bibinfo {pages} {3426} (\bibinfo {year}
  {1997})}\BibitemShut {NoStop}%
\bibitem [{\citenamefont {{Henriksen}}\ and\ \citenamefont
  {{Widrow}}(1999)}]{Henriksen1999}%
  \BibitemOpen
  \bibfield  {author} {\bibinfo {author} {\bibfnamefont {R.~N.}\ \bibnamefont
  {{Henriksen}}}\ and\ \bibinfo {author} {\bibfnamefont {L.~M.}\ \bibnamefont
  {{Widrow}}},\ }\href {\doibase 10.1046/j.1365-8711.1999.02124.x} {\bibfield
  {journal} {\bibinfo  {journal} {\mnras}\ }\textbf {\bibinfo {volume} {302}},\
  \bibinfo {pages} {321} (\bibinfo {year} {1999})},\ \Eprint
  {http://arxiv.org/abs/astro-ph/9805277} {astro-ph/9805277} \BibitemShut
  {NoStop}%
\bibitem [{\citenamefont {{Riebe}}\ \emph {et~al.}(2013)\citenamefont
  {{Riebe}}, \citenamefont {{Partl}}, \citenamefont {{Enke}}, \citenamefont
  {{Forero-Romero}}, \citenamefont {{Gottl{\"o}ber}}, \citenamefont {{Klypin}},
  \citenamefont {{Lemson}}, \citenamefont {{Prada}}, \citenamefont {{Primack}},
  \citenamefont {{Steinmetz}},\ and\ \citenamefont
  {{Turchaninov}}}]{Multidark}%
  \BibitemOpen
  \bibfield  {author} {\bibinfo {author} {\bibfnamefont {K.}~\bibnamefont
  {{Riebe}}}, \bibinfo {author} {\bibfnamefont {A.~M.}\ \bibnamefont
  {{Partl}}}, \bibinfo {author} {\bibfnamefont {H.}~\bibnamefont {{Enke}}},
  \bibinfo {author} {\bibfnamefont {J.}~\bibnamefont {{Forero-Romero}}},
  \bibinfo {author} {\bibfnamefont {S.}~\bibnamefont {{Gottl{\"o}ber}}},
  \bibinfo {author} {\bibfnamefont {A.}~\bibnamefont {{Klypin}}}, \bibinfo
  {author} {\bibfnamefont {G.}~\bibnamefont {{Lemson}}}, \bibinfo {author}
  {\bibfnamefont {F.}~\bibnamefont {{Prada}}}, \bibinfo {author} {\bibfnamefont
  {J.~R.}\ \bibnamefont {{Primack}}}, \bibinfo {author} {\bibfnamefont
  {M.}~\bibnamefont {{Steinmetz}}}, \ and\ \bibinfo {author} {\bibfnamefont
  {V.}~\bibnamefont {{Turchaninov}}},\ }\href {\doibase 10.1002/asna.201211900}
  {\bibfield  {journal} {\bibinfo  {journal} {Astronomische Nachrichten}\
  }\textbf {\bibinfo {volume} {334}},\ \bibinfo {pages} {691} (\bibinfo {year}
  {2013})}\BibitemShut {NoStop}%
\bibitem [{\citenamefont {{Prada}}\ \emph {et~al.}(2012)\citenamefont
  {{Prada}}, \citenamefont {{Klypin}}, \citenamefont {{Cuesta}}, \citenamefont
  {{Betancort-Rijo}},\ and\ \citenamefont {{Primack}}}]{Prada2012}%
  \BibitemOpen
  \bibfield  {author} {\bibinfo {author} {\bibfnamefont {F.}~\bibnamefont
  {{Prada}}}, \bibinfo {author} {\bibfnamefont {A.~A.}\ \bibnamefont
  {{Klypin}}}, \bibinfo {author} {\bibfnamefont {A.~J.}\ \bibnamefont
  {{Cuesta}}}, \bibinfo {author} {\bibfnamefont {J.~E.}\ \bibnamefont
  {{Betancort-Rijo}}}, \ and\ \bibinfo {author} {\bibfnamefont
  {J.}~\bibnamefont {{Primack}}},\ }\href {\doibase
  10.1111/j.1365-2966.2012.21007.x} {\bibfield  {journal} {\bibinfo  {journal}
  {\mnras}\ }\textbf {\bibinfo {volume} {423}},\ \bibinfo {pages} {3018}
  (\bibinfo {year} {2012})},\ \Eprint {http://arxiv.org/abs/1104.5130}
  {arXiv:1104.5130 [astro-ph.CO]} \BibitemShut {NoStop}%
\bibitem [{\citenamefont {{Klypin}}\ \emph {et~al.}(2011)\citenamefont
  {{Klypin}}, \citenamefont {{Trujillo-Gomez}},\ and\ \citenamefont
  {{Primack}}}]{Klypin2011}%
  \BibitemOpen
  \bibfield  {author} {\bibinfo {author} {\bibfnamefont {A.~A.}\ \bibnamefont
  {{Klypin}}}, \bibinfo {author} {\bibfnamefont {S.}~\bibnamefont
  {{Trujillo-Gomez}}}, \ and\ \bibinfo {author} {\bibfnamefont
  {J.}~\bibnamefont {{Primack}}},\ }\href {\doibase
  10.1088/0004-637X/740/2/102} {\bibfield  {journal} {\bibinfo  {journal}
  {\apj}\ }\textbf {\bibinfo {volume} {740}},\ \bibinfo {eid} {102} (\bibinfo
  {year} {2011})},\ \Eprint {http://arxiv.org/abs/1002.3660} {arXiv:1002.3660
  [astro-ph.CO]} \BibitemShut {NoStop}%
\bibitem [{\citenamefont {{Behroozi}}\ \emph
  {et~al.}(2013{\natexlab{a}})\citenamefont {{Behroozi}}, \citenamefont
  {{Wechsler}},\ and\ \citenamefont {{Wu}}}]{Rockstar}%
  \BibitemOpen
  \bibfield  {author} {\bibinfo {author} {\bibfnamefont {P.~S.}\ \bibnamefont
  {{Behroozi}}}, \bibinfo {author} {\bibfnamefont {R.~H.}\ \bibnamefont
  {{Wechsler}}}, \ and\ \bibinfo {author} {\bibfnamefont {H.-Y.}\ \bibnamefont
  {{Wu}}},\ }\href {\doibase 10.1088/0004-637X/762/2/109} {\bibfield  {journal}
  {\bibinfo  {journal} {\apj}\ }\textbf {\bibinfo {volume} {762}},\ \bibinfo
  {eid} {109} (\bibinfo {year} {2013}{\natexlab{a}})},\ \Eprint
  {http://arxiv.org/abs/1110.4372} {arXiv:1110.4372 [astro-ph.CO]} \BibitemShut
  {NoStop}%
\bibitem [{\citenamefont {{Behroozi}}\ \emph
  {et~al.}(2013{\natexlab{b}})\citenamefont {{Behroozi}}, \citenamefont
  {{Wechsler}}, \citenamefont {{Wu}}, \citenamefont {{Busha}}, \citenamefont
  {{Klypin}},\ and\ \citenamefont {{Primack}}}]{Behroozi2013}%
  \BibitemOpen
  \bibfield  {author} {\bibinfo {author} {\bibfnamefont {P.~S.}\ \bibnamefont
  {{Behroozi}}}, \bibinfo {author} {\bibfnamefont {R.~H.}\ \bibnamefont
  {{Wechsler}}}, \bibinfo {author} {\bibfnamefont {H.-Y.}\ \bibnamefont
  {{Wu}}}, \bibinfo {author} {\bibfnamefont {M.~T.}\ \bibnamefont {{Busha}}},
  \bibinfo {author} {\bibfnamefont {A.~A.}\ \bibnamefont {{Klypin}}}, \ and\
  \bibinfo {author} {\bibfnamefont {J.~R.}\ \bibnamefont {{Primack}}},\ }\href
  {\doibase 10.1088/0004-637X/763/1/18} {\bibfield  {journal} {\bibinfo
  {journal} {\apj}\ }\textbf {\bibinfo {volume} {763}},\ \bibinfo {eid} {18}
  (\bibinfo {year} {2013}{\natexlab{b}})},\ \Eprint
  {http://arxiv.org/abs/1110.4370} {arXiv:1110.4370 [astro-ph.CO]} \BibitemShut
  {NoStop}%
\bibitem [{Note2()}]{Note2}%
  \BibitemOpen
  \bibinfo {note} {http://hipacc.ucsc.edu/Bolshoi/MergerTrees.html}\BibitemShut
  {NoStop}%
\bibitem [{\citenamefont {{Wechsler}}\ \emph {et~al.}(2002)\citenamefont
  {{Wechsler}}, \citenamefont {{Bullock}}, \citenamefont {{Primack}},
  \citenamefont {{Kravtsov}},\ and\ \citenamefont {{Dekel}}}]{Wechsler2002}%
  \BibitemOpen
  \bibfield  {author} {\bibinfo {author} {\bibfnamefont {R.~H.}\ \bibnamefont
  {{Wechsler}}}, \bibinfo {author} {\bibfnamefont {J.~S.}\ \bibnamefont
  {{Bullock}}}, \bibinfo {author} {\bibfnamefont {J.~R.}\ \bibnamefont
  {{Primack}}}, \bibinfo {author} {\bibfnamefont {A.~V.}\ \bibnamefont
  {{Kravtsov}}}, \ and\ \bibinfo {author} {\bibfnamefont {A.}~\bibnamefont
  {{Dekel}}},\ }\href {\doibase 10.1086/338765} {\bibfield  {journal} {\bibinfo
   {journal} {\apj}\ }\textbf {\bibinfo {volume} {568}},\ \bibinfo {pages} {52}
  (\bibinfo {year} {2002})},\ \Eprint {http://arxiv.org/abs/astro-ph/0108151}
  {astro-ph/0108151} \BibitemShut {NoStop}%
\bibitem [{\citenamefont {{Diemer}}\ \emph {et~al.}(2013)\citenamefont
  {{Diemer}}, \citenamefont {{More}},\ and\ \citenamefont
  {{Kravtsov}}}]{Diemer2013}%
  \BibitemOpen
  \bibfield  {author} {\bibinfo {author} {\bibfnamefont {B.}~\bibnamefont
  {{Diemer}}}, \bibinfo {author} {\bibfnamefont {S.}~\bibnamefont {{More}}}, \
  and\ \bibinfo {author} {\bibfnamefont {A.~V.}\ \bibnamefont {{Kravtsov}}},\
  }\href {\doibase 10.1088/0004-637X/766/1/25} {\bibfield  {journal} {\bibinfo
  {journal} {\apj}\ }\textbf {\bibinfo {volume} {766}},\ \bibinfo {eid} {25}
  (\bibinfo {year} {2013})},\ \Eprint {http://arxiv.org/abs/1207.0816}
  {arXiv:1207.0816 [astro-ph.CO]} \BibitemShut {NoStop}%
\bibitem [{\citenamefont {{Bardeen}}\ \emph {et~al.}(1986)\citenamefont
  {{Bardeen}}, \citenamefont {{Bond}}, \citenamefont {{Kaiser}},\ and\
  \citenamefont {{Szalay}}}]{BBKS}%
  \BibitemOpen
  \bibfield  {author} {\bibinfo {author} {\bibfnamefont {J.~M.}\ \bibnamefont
  {{Bardeen}}}, \bibinfo {author} {\bibfnamefont {J.~R.}\ \bibnamefont
  {{Bond}}}, \bibinfo {author} {\bibfnamefont {N.}~\bibnamefont {{Kaiser}}}, \
  and\ \bibinfo {author} {\bibfnamefont {A.~S.}\ \bibnamefont {{Szalay}}},\
  }\href {\doibase 10.1086/164143} {\bibfield  {journal} {\bibinfo  {journal}
  {\apj}\ }\textbf {\bibinfo {volume} {304}},\ \bibinfo {pages} {15} (\bibinfo
  {year} {1986})}\BibitemShut {NoStop}%
\bibitem [{\citenamefont {{Gill}}\ \emph {et~al.}(2005)\citenamefont {{Gill}},
  \citenamefont {{Knebe}},\ and\ \citenamefont {{Gibson}}}]{Gill2005}%
  \BibitemOpen
  \bibfield  {author} {\bibinfo {author} {\bibfnamefont {S.~P.~D.}\
  \bibnamefont {{Gill}}}, \bibinfo {author} {\bibfnamefont {A.}~\bibnamefont
  {{Knebe}}}, \ and\ \bibinfo {author} {\bibfnamefont {B.~K.}\ \bibnamefont
  {{Gibson}}},\ }\href {\doibase 10.1111/j.1365-2966.2004.08562.x} {\bibfield
  {journal} {\bibinfo  {journal} {\mnras}\ }\textbf {\bibinfo {volume} {356}},\
  \bibinfo {pages} {1327} (\bibinfo {year} {2005})},\ \Eprint
  {http://arxiv.org/abs/astro-ph/0404427} {astro-ph/0404427} \BibitemShut
  {NoStop}%
\bibitem [{\citenamefont {{Ludlow}}\ \emph {et~al.}(2009)\citenamefont
  {{Ludlow}}, \citenamefont {{Navarro}}, \citenamefont {{Springel}},
  \citenamefont {{Jenkins}}, \citenamefont {{Frenk}},\ and\ \citenamefont
  {{Helmi}}}]{Ludlow2009}%
  \BibitemOpen
  \bibfield  {author} {\bibinfo {author} {\bibfnamefont {A.~D.}\ \bibnamefont
  {{Ludlow}}}, \bibinfo {author} {\bibfnamefont {J.~F.}\ \bibnamefont
  {{Navarro}}}, \bibinfo {author} {\bibfnamefont {V.}~\bibnamefont
  {{Springel}}}, \bibinfo {author} {\bibfnamefont {A.}~\bibnamefont
  {{Jenkins}}}, \bibinfo {author} {\bibfnamefont {C.~S.}\ \bibnamefont
  {{Frenk}}}, \ and\ \bibinfo {author} {\bibfnamefont {A.}~\bibnamefont
  {{Helmi}}},\ }\href {\doibase 10.1088/0004-637X/692/1/931} {\bibfield
  {journal} {\bibinfo  {journal} {\apj}\ }\textbf {\bibinfo {volume} {692}},\
  \bibinfo {pages} {931} (\bibinfo {year} {2009})},\ \Eprint
  {http://arxiv.org/abs/0801.1127} {arXiv:0801.1127} \BibitemShut {NoStop}%
\bibitem [{\citenamefont {{Anderhalden}}\ and\ \citenamefont
  {{Diemand}}(2011)}]{Anderhalden2011}%
  \BibitemOpen
  \bibfield  {author} {\bibinfo {author} {\bibfnamefont {D.}~\bibnamefont
  {{Anderhalden}}}\ and\ \bibinfo {author} {\bibfnamefont {J.}~\bibnamefont
  {{Diemand}}},\ }\href {\doibase 10.1111/j.1365-2966.2011.18614.x} {\bibfield
  {journal} {\bibinfo  {journal} {\mnras}\ }\textbf {\bibinfo {volume} {414}},\
  \bibinfo {pages} {3166} (\bibinfo {year} {2011})},\ \Eprint
  {http://arxiv.org/abs/1102.5736} {arXiv:1102.5736 [astro-ph.CO]} \BibitemShut
  {NoStop}%
\bibitem [{\citenamefont {{Wetzel}}\ \emph {et~al.}(2014)\citenamefont
  {{Wetzel}}, \citenamefont {{Tinker}}, \citenamefont {{Conroy}},\ and\
  \citenamefont {{Bosch}}}]{Wetzel2014}%
  \BibitemOpen
  \bibfield  {author} {\bibinfo {author} {\bibfnamefont {A.~R.}\ \bibnamefont
  {{Wetzel}}}, \bibinfo {author} {\bibfnamefont {J.~L.}\ \bibnamefont
  {{Tinker}}}, \bibinfo {author} {\bibfnamefont {C.}~\bibnamefont {{Conroy}}},
  \ and\ \bibinfo {author} {\bibfnamefont {F.~C.~v.~d.}\ \bibnamefont
  {{Bosch}}},\ }\href {\doibase 10.1093/mnras/stu122} {\bibfield  {journal}
  {\bibinfo  {journal} {\mnras}\ }\textbf {\bibinfo {volume} {439}},\ \bibinfo
  {pages} {2687} (\bibinfo {year} {2014})},\ \Eprint
  {http://arxiv.org/abs/1303.7231} {arXiv:1303.7231 [astro-ph.CO]} \BibitemShut
  {NoStop}%
\bibitem [{\citenamefont {{Garrison-Kimmel}}\ \emph {et~al.}(2014)\citenamefont
  {{Garrison-Kimmel}}, \citenamefont {{Boylan-Kolchin}}, \citenamefont
  {{Bullock}},\ and\ \citenamefont {{Lee}}}]{Garrison-Kimmel2014}%
  \BibitemOpen
  \bibfield  {author} {\bibinfo {author} {\bibfnamefont {S.}~\bibnamefont
  {{Garrison-Kimmel}}}, \bibinfo {author} {\bibfnamefont {M.}~\bibnamefont
  {{Boylan-Kolchin}}}, \bibinfo {author} {\bibfnamefont {J.~S.}\ \bibnamefont
  {{Bullock}}}, \ and\ \bibinfo {author} {\bibfnamefont {K.}~\bibnamefont
  {{Lee}}},\ }\href {\doibase 10.1093/mnras/stt2377} {\bibfield  {journal}
  {\bibinfo  {journal} {\mnras}\ }\textbf {\bibinfo {volume} {438}},\ \bibinfo
  {pages} {2578} (\bibinfo {year} {2014})},\ \Eprint
  {http://arxiv.org/abs/1310.6746} {arXiv:1310.6746 [astro-ph.CO]} \BibitemShut
  {NoStop}%
\bibitem [{\citenamefont {{Oguri}}\ and\ \citenamefont
  {{Hamana}}(2011)}]{Oguri2011}%
  \BibitemOpen
  \bibfield  {author} {\bibinfo {author} {\bibfnamefont {M.}~\bibnamefont
  {{Oguri}}}\ and\ \bibinfo {author} {\bibfnamefont {T.}~\bibnamefont
  {{Hamana}}},\ }\href {\doibase 10.1111/j.1365-2966.2011.18481.x} {\bibfield
  {journal} {\bibinfo  {journal} {\mnras}\ }\textbf {\bibinfo {volume} {414}},\
  \bibinfo {pages} {1851} (\bibinfo {year} {2011})},\ \Eprint
  {http://arxiv.org/abs/1101.0650} {arXiv:1101.0650 [astro-ph.CO]} \BibitemShut
  {NoStop}%
\bibitem [{\citenamefont {{Dalal}}\ \emph {et~al.}(2008)\citenamefont
  {{Dalal}}, \citenamefont {{White}}, \citenamefont {{Bond}},\ and\
  \citenamefont {{Shirokov}}}]{Dalal08}%
  \BibitemOpen
  \bibfield  {author} {\bibinfo {author} {\bibfnamefont {N.}~\bibnamefont
  {{Dalal}}}, \bibinfo {author} {\bibfnamefont {M.}~\bibnamefont {{White}}},
  \bibinfo {author} {\bibfnamefont {J.~R.}\ \bibnamefont {{Bond}}}, \ and\
  \bibinfo {author} {\bibfnamefont {A.}~\bibnamefont {{Shirokov}}},\ }\href
  {\doibase 10.1086/591512} {\bibfield  {journal} {\bibinfo  {journal} {\apj}\
  }\textbf {\bibinfo {volume} {687}},\ \bibinfo {pages} {12} (\bibinfo {year}
  {2008})},\ \Eprint {http://arxiv.org/abs/arXiv:0803.3453}
  {arXiv:arXiv:0803.3453} \BibitemShut {NoStop}%
\bibitem [{\citenamefont {{Lau}}\ \emph {et~al.}(2014)\citenamefont {{Lau}},
  \citenamefont {{Nagai}},\ and\ \citenamefont {{Nelson}}}]{Lau2014}%
  \BibitemOpen
  \bibfield  {author} {\bibinfo {author} {\bibfnamefont {E.}~\bibnamefont
  {{Lau}}}, \bibinfo {author} {\bibfnamefont {D.}~\bibnamefont {{Nagai}}}, \
  and\ \bibinfo {author} {\bibfnamefont {K.}~\bibnamefont {{Nelson}}},\
  }\href@noop {} {\bibfield  {journal} {\bibinfo  {journal} {in preparation}\ }
  (\bibinfo {year} {2014})}\BibitemShut {NoStop}%
\end{thebibliography}

%merlin.mbs apsrev4-1.bst 2010-07-25 4.21a (PWD, AO, DPC) hacked
%Control: key (0)
%Control: author (8) initials jnrlst
%Control: editor formatted (1) identically to author
%Control: production of article title (-1) disabled
%Control: page (0) single
%Control: year (1) truncated
%Control: production of eprint (0) enabled
%

\end{document}